\documentclass[aps,prl,11pt,onecolumn,superscriptaddress]{revtex4-2}
\usepackage{graphicx}
\usepackage{amsmath,amsfonts}
\usepackage{color}
\usepackage{ulem}
\usepackage{braket}
\newcommand{\upsub}[1]{\sb{\mathrm{#1}}}
\newcommand{\upsup}[1]{\sp{\mathrm{#1}}}

\begingroup\lccode`~=`_\lowercase{\endgroup\let~\upsub}
\begingroup\lccode`~=`^\lowercase{\endgroup\let~\upsup}

\AtBeginDocument{%
	\catcode`_=12 \catcode`^=12
	\mathcode`_="8000
	\mathcode`^="8000
}

\begin{document}


	\title{Radiative suppression of exciton-exciton annihilation in a two-dimensional semiconductor}

	\author{Luca Sortino}
	\email{luca.sortino@lmu.de}
		\affiliation{Chair in Hybrid Nanosystems, Nanoinstitute Munich, Faculty of Physics, Ludwig-Maximilians-Universit{\"a}t M{\"u}nchen, 80539 Munich, Germany}
		\affiliation{Center for NanoScience, Faculty of Physics, Ludwig-Maximilians-Universit{\"a}t M{\"u}nchen, 80539 Munich, Germany}
	\author{Merve G{\"u}lm{\"u}s}
			\affiliation{Chair in Hybrid Nanosystems, Nanoinstitute Munich, Faculty of Physics, Ludwig-Maximilians-Universit{\"a}t M{\"u}nchen, 80539 Munich, Germany}
	\author{Benjamin Tilmann}
	\affiliation{Chair in Hybrid Nanosystems, Nanoinstitute Munich, Faculty of Physics, Ludwig-Maximilians-Universit{\"a}t M{\"u}nchen, 80539 Munich, Germany}
	\affiliation{Center for NanoScience, Faculty of Physics, Ludwig-Maximilians-Universit{\"a}t M{\"u}nchen, 80539 Munich, Germany}
	
	\author{Leonardo de S. Menezes}
\affiliation{Chair in Hybrid Nanosystems, Nanoinstitute Munich, Faculty of Physics, Ludwig-Maximilians-Universit{\"a}t M{\"u}nchen, 80539 Munich, Germany}
\affiliation{Center for NanoScience, Faculty of Physics, Ludwig-Maximilians-Universit{\"a}t M{\"u}nchen, 80539 Munich, Germany}

	\affiliation{Departamento de F\'{i}sica, Universidade Federal de Pernambuco, 50670-901 Recife-PE, Brazil}
	\author{Stefan A. Maier}
	\affiliation{School of Physics and Astronomy, Monash University, Clayton, Victoria 3800, Australia}
	\affiliation{The Blackett Laboratory, Department of Physics, Imperial College London, London, SW7 2BW, United Kingdom}
\affiliation{Chair in Hybrid Nanosystems, Nanoinstitute Munich, Faculty of Physics, Ludwig-Maximilians-Universit{\"a}t M{\"u}nchen, 80539 Munich, Germany}
\affiliation{Center for NanoScience, Faculty of Physics, Ludwig-Maximilians-Universit{\"a}t M{\"u}nchen, 80539 Munich, Germany}

	\date{\today}
	\maketitle

\noindent
\textbf{Two-dimensional (2D) semiconductors possess strongly bound excitons, opening novel opportunities for engineering light-matter interaction at the nanoscale. 
However, their in-plane confinement leads to large non-radiative exciton-exciton annihilation (EEA) processes, setting a fundamental limit for their photonic applications.
In this work, we demonstrate suppression of EEA via enhancement of light-matter interaction in hybrid 2D semiconductor-dielectric nanophotonic platforms, by coupling excitons in WS$ _2 $ monolayers with optical Mie resonances in dielectric nanoantennas. The hybrid system reaches an intermediate light-matter coupling regime, with photoluminescence enhancement factors up to 10$ ^2 $. 
Probing the exciton ultrafast dynamics reveal suppressed EEA for coupled excitons, even under high exciton densities $>$ 10$^{12}$ cm$^{-2} $. We extract EEA coefficients in the order of 10$^{-3} $, compared to 10$^{-2} $ for uncoupled monolayers, as well as absorption enhancement of 3.9 and a Purcell factor of 4.5. Our results highlight engineering the photonic environment as a route to achieve higher quantum efficiencies for low-power hybrid devices, and larger exciton densities, towards strongly correlated excitonic phases in 2D semiconductors.
}

\newpage
\pagebreak

\noindent
The Auger-Meitner effect in semiconductor is a scattering process where two charge carriers collide, resulting in a non-radiative decay via a mutual exchange of momentum \cite{Klingshirn2012a}. 
It represents a major loss channel in optoelectronic devices, posing a fundamental limit on their quantum efficiency under high carrier densities. 
In particular, for low-dimensional semiconductors, quantum confinement restricts the momentum conservation rules of carriers and allows for stable Coulomb bound electron-hole pairs, or excitons. 
These can scatter via the mutual dipole-dipole interaction, in the form of exciton-exciton annihilation (EEA), yielding large scattering rates compared to bulk materials, as observed in quantum dots \cite{Efros2016}, quantum wells \cite{Shen2007} and carbon nanotubes \cite{Wang2004}.
In the case of two-dimensional (2D) semiconductors, control and suppression of EEA is fundamental to unlock their potential for applications \cite{Yu2016a}.
Transition metal dichalcogenides (TMDCs) emerged as the most promising family of atomically thin semiconductors for photonic applications \cite{Mak2016}. Owing to large exciton binding energies above 200 meV, TMDCs optical properties are dominated by their excitonic response up to room temperature \cite{Wang2017}, while at cryogenic temperatures TMDCs exhibit appealing properties, such as the presence of many body excitonic species \cite{Rodin2020} or single photon emitters \cite{Turunen2022}. 
However, excitons in 2D TMDCs possess large Bohr radii, in the order of 1 nm, increasing their mutual interaction ranges and resulting in large EEA coefficients, reaching values larger than in any other semiconducting material \cite{Yu2016a}. EEA is thus observed even at relatively low exciton populations, setting a fundamental limit for the generation of high exciton densities in 2D semiconductors.  
Experimental techniques, such as time resolved luminescence or ultrafast transient absorption spectroscopy, have been employed to study EEA processes in atomically thin and bulk TMDCs \cite{Kumar2014a,Sun2014, Yuan2015b, Mouri2014}. EEA effect introduces a drastic change the exciton dynamics, observed as a fast recombination process, in the order of few picoseconds, which follows a quadratic dependence with the generated exciton population \cite{Sun2014}.
Recent works explored various approaches for suppressing EEA processes. For instance, by encapsulating TMDCs monolayers in hexagonal boron nitride \cite{Masubuchi2017}, extracting excess free carriers \cite{Lien2019}, or the simultaneous application of strain and a gate voltage \cite{Kim2021}. 
Notably, EEA processes in 2D semiconductors can also be harnessed to provide unexpected effects, for instance in generation of upconverted photoluminescence \cite{Han2018,Binder2019}, increased photocurrents \cite{Linardy2020a}, and creation of negative mass excitons \cite{Lin2021}.

An alternative approach to overcome the limitations imposed by EEA is offered by the integration of 2D semiconductors in nanophotonic architectures,  tailoring the dielectric environment and the local density of states experienced by 2D confined excitons. The ability of 2D TMDCs to conform to underlying nanophotonic structures, and couple to the strong near field at their surfaces, have been demonstrated to enhance light-matter interaction in excitons \cite{Sortino2019,Yuan2021a,Fang2022,Zotev2022,Petric2022} and single photon emitters \cite{Luo2018,Sortino2021a}, making them a promising material for hybrid nanophotonic devices. 
Resonant dielectric optical nanoantennas recently emerged as a novel platform to overcome the intrinsic losses of metal based plasmonic counterparts, while providing a new toolbox to tailor light-matter interaction at the nanoscale \cite{Kuznetsov2016}. By sustaining the presence of both electric and magnetic type of optical resonances, multimodal interference of electromagnetic Mie modes in a single dielectric nanoantenna opens to higher degrees of control on light-matter interaction, from unidirectional scattering effects \cite{Staude2013a} to suppression of far field emission \cite{Miroshnichenko2015}. This approach can be further extended to arrays of nanoantennas, or metasurfaces, for the manipulation of phase and amplitude of light in sub-wavelength dimensions and the physics of bound states in the continuum \cite{Cortes2022}.

\begin{figure*}
	\centering
	\includegraphics[width=1\linewidth]{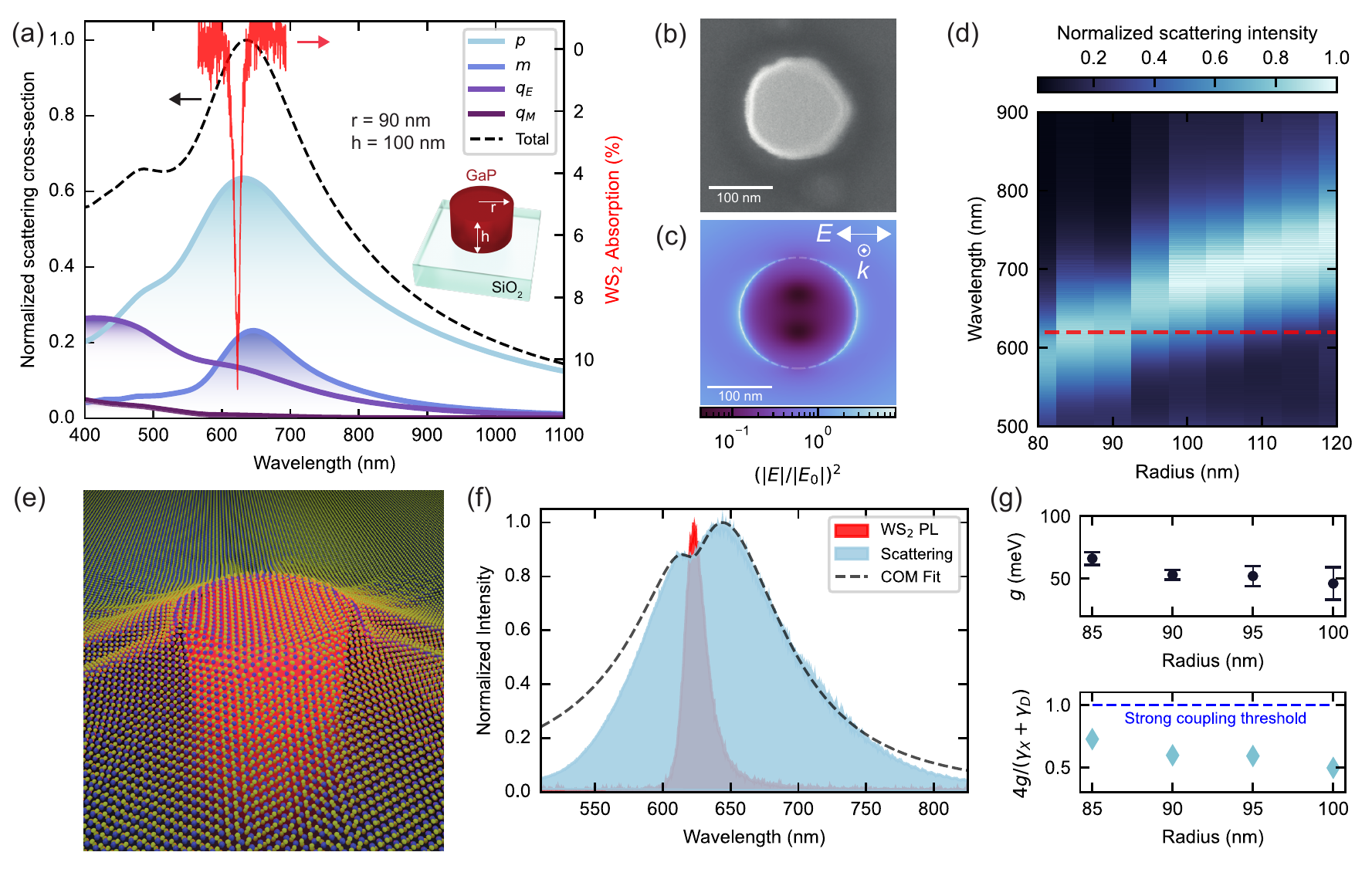}
	\caption{\textbf{Hybrid 2D semiconductor-dielectric nanoantennas for enhanced light-matter coupling regime in 2D TMDCs} (a) Simulated scattering cross section  of a GaP nanoantenna with radius of 90 nm and height of 100 nm (dashed black line), and absorption of the \textit{1s}-exciton state of a WS$ _2 $ monolayer on SiO$ _2 $ substrate (red line). The scattering cross section is analysed with a multipolar expansion, where the respective weights of the electrical ($ p $) and magnetic ($ m $) dipoles, and the electrical ($ q_E $) and magnetic ($ q_M $) quadrupoles are shown. (b) Electron microscope image of a fabricated GaP nanoantenna. (c) FDTD simulated electric field intensity, $(|E|/|E_0|)^{2}$, at the top surface of a GaP nanoantenna (r = 90 nm, h = 100 nm) at $\lambda=530$ nm. (d) Dark field scattering spectra of fabricated GaP nanoantennas for different radial dimensions. The dashed red line at 620 nm represents the resonance wavelength of WS$ _2 $ monolayer excitons. (e) Illustration of the hybrid nanophotonic system composed of a WS$ _2 $ monolayer coupled to a cylindrical GaP dielectric nanoantenna. (f) PL of nanoantenna coupled WS$ _2 $ (in red) and dark field scattering (in light blue) of the GaP nanoantennas (radius 90 nm, height 100 nm). The scattering spectra is fitted with a coupled oscillator model (COM - dashed black line). (g) Upper panel: coupling strength ($ g $) values extracted from the coupled oscillator model used to fit the scattering spectra of nanoantennas coupled with WS$_2$ monolayers. Lower panel: Ratio of the extracted coupling strengths values compared with the normalized strong coupling condition $ 4g/(\gamma_X+\gamma_D)=1$.}
	\label{fig:figure2dantenna}
\end{figure*}

In this work, we demonstrate the suppression of EEA processes via radiative rate enhancement, by coupling excitons in WS$ _2 $ monolayers with Mie resonances of gallium phosphide (GaP) dielectric nanoantennas \cite{Cambiasso2017a}. We show that the hybrid 2D semiconductor-dielectric nanoantenna system reaches an intermediate light-matter coupling regime and observe photoluminescence (PL) enhancement factors above 10$ ^2 $ compared to uncoupled monolayers, as well as a reduction of the PL lifetime, a signature of spontaneous emission rate enhancement. We then probe the exciton dynamics with ultrafast transient absorption spectroscopy. For uncoupled excitons, we observe the expected onset of non-radiative EEA as a fast recombination process in their dynamics \cite{Sun2014}. On the contrary, excitons coupled to the near fields of GaP nanoantennas exhibit negligible changes in their dynamics over a broad range of excitation powers. 
We show this is the combined effect of the enhanced absorption rate via near field coupling, and increased spontaneous emission rate via the Purcell effect, leading to excitons having higher probability to recombine radiatively, rather than experiencing diffusion and non-radiative processes. 
In the framework of a rate equation model, we extract the values of the EEA coefficient ($ k_{A}$) and found one order of magnitude lower values for WS$ _2 $ excitons coupled to resonant nanoantennas, as compared to uncoupled excitons on glass substrate.
Moreover, by comparing their ultrafast dynamics, we extract an enhanced absorption rate ($ \sigma $) of 3.9 and a Purcell factor ($ F_P $) of 4.5. 
This behaviour goes against the phenomenological law of decreasing EEA coefficients with longer exciton lifetimes \cite{Uddin2021}, highlighting enhanced light-matter interaction as a key for the suppression of EEA processes in 2D semiconductors.
Our results demonstrate hybrid nanophotonics architectures as an attractive platform to engineer light-matter coupling with 2D materials and provide a route to overcome fundamental limitations induced by exciton scattering, enabling application of 2D semiconductors in photonic devices.

\noindent
\textbf{Results}

\noindent
\textbf{Intermediate light-matter coupling regime in hybrid 2D semiconductor-dielectric nanoantennas} We select the geometry of the GaP nanoantenna to maximize the spectral overlap between the magnetic and electric dipolar Mie resonances and excitons in WS$ _2 $ monolayers. 
Figure \ref{fig:figure2dantenna}a shows the finite-difference time-domain (FDTD) numerical simulation of the scattering spectrum for a single GaP nanoantenna on a SiO$ _2 $ glass substrate (dashed black line), with a radius of 90 nm and height of 100 nm. The red curve represents the experimental optical absorption of the \textit{1s}-exciton state in a monolayer WS$ _2 $ on SiO$ _2 $ substrate, also referred to as A exciton in literature \cite{Wang2017}. The scattering cross-section can be described with a multipolar expansion of the induced electromagnetic currents \cite{Cortes2022}, quantifying the individual contributions from the optical Mie resonances, respectively the electrical ($ p $) and magnetic ($ m $) dipoles, and the electrical ($ q_E $) and magnetic ($ q_M $) quadrupoles. 
We fabricated an array of optical nanoantennas by depositing thermally grown amorphous GaP on top of fused silica substrates, and patterned the thin film with conventional electron beam lithography and reactive ion etching techniques (see Methods for details). Figure \ref{fig:figure2dantenna}b shows a top view electron microscope image of a fabricated cylindrical GaP nanoantenna on SiO$ _2 $ substrate. 
At its surface, the nanoantenna confines and enhances the electromagnetic field intensity, $(|E|/|E_{0}|)^2$, defined as the ratio between the electric field amplitude of the scattered field by the antenna ($ E $) and the normally incident field ($ E_0 $). Figure \ref{fig:figure2dantenna}c shows the numerical simulations for the enhanced near field region, recorded at the top surface of a single GaP nanoantenna. 
By tuning the radial dimension of the nanoantennas, we tailor the wavelength of the Mie resonances to match with the WS$ _2 $ exciton wavelength. Figure \ref{fig:figure2dantenna}d shows the darkfield scattering spectra of the fabricated GaP nanoantennas array, in good agreement with the numerical simulations in Figure \ref{fig:figure2dantenna}a. As expected from Mie theory, increasing the resonator size shifts the Mie resonances to lower energies, crossing the WS$_2$ exciton energy (dashed red line in Figure \ref{fig:figure2dantenna}d).

To probe the coupled system, we transfer the WS$_2$ monolayer on top of the nanoantenna array with an all-dry transfer technique (see Methods and Supplementary Note I). Figure \ref{fig:figure2dantenna}e displays an illustration of a monolayer WS$ _2 $ transferred on top of a cylindric GaP dielectric nanoantenna on a glass substrate. The atomically thin layer stretches on top of the nanoantenna, in close proximity with the enhanced near field region. Figure \ref{fig:figure2dantenna}f shows the WS$_2$ PL emission and the dark field scattering spectra of a hybrid nanoantenna covered with a WS$ _2 $ monolayer. The scattering spectrum is modified by the presence of the atomically thin layer, in the form of a dip in correspondence to the PL exciton peak of the coupled WS$_2$ monolayer (see also Supplementary Note II), indicating an enhanced absorption via the resonant coupling between excitons and Mie resonances \cite{Pelton2019}.
We treat the nanoantennas optical resonances and WS$ _2 $ excitons as damped coupled oscillators, and fit the scattering spectrum of the hybrid system with a coupled oscillator model (COM) in the form \cite{Pelton2019}: 

\begin{equation}
	\sigma_{scatt}(\omega) = A\omega^4  \left| \dfrac{ (\omega_X^2-\omega^2-i\omega\gamma_X)}{(\omega_D^2-\omega^2-i\omega\gamma_D)(\omega_X^2-\omega^2-i\omega\gamma_X)-\omega_X\omega_D g^2} \right| ^2
\end{equation}

\smallskip

\noindent
where $ \gamma_X $ and $ \gamma_D $ are the exciton and antenna dipolar resonance linewidths, respectively, $ \omega_X $ and $ \omega_D $ the exciton and antenna resonance frequencies, $ A $ is a scaling constant, and $ g $ the coupling strength constant.
 In Figure \ref{fig:figure2dantenna}g we plot the extracted values of $ g $ (top panel) for WS$ _2 $ coupled to different antennas with radius ranging from 85 nm to 100 nm. We found values in the range of 50-60 meV, comparable with similar hybrid architectures based on plasmonic nanoantennas \cite{Petric2022}.
 We then compare the extracted values of $ g $ with the strong coupling condition satisfying that $2g > \frac{1}{2} (\gamma_X+\gamma_D)$ \cite{Pelton2019} (Figure \ref{fig:figure2dantenna}g, bottom panel). Due to the variation in linewidth of the fabricated antennas and the strain affecting the exciton resonance linewidth, we define a normalized value of the strong coupling condition as $ 4g/(\gamma_X+\gamma_D)=1$. For all the hybrid  systems studied, we obtain values where $ g > 0.5 $, confirming the increased light-matter interaction of WS$ _2 $ excitons coupled to optical Mie resonances, and placing our hybrid 2D semiconductor-dielectric system in the intermediate light-matter coupling regime. 

\begin{figure*}[t]
	\centering
	\includegraphics[width=1\linewidth]{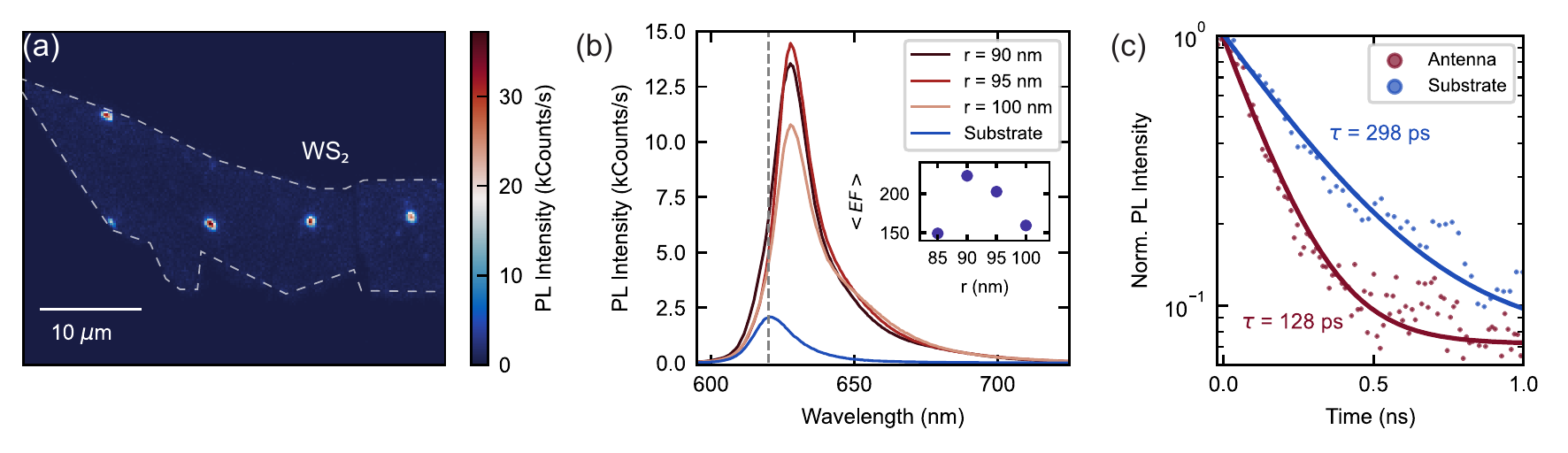}
	\caption{\textbf{Photoluminescence enhancement and Purcell effect in WS$ _{2} $ monolayers coupled to GaP nanoantennas} (a) Map of WS$ _2 $ monolayer PL on top of an array of resonant GaP nanoantennas. (b) PL signal of the WS$ _2 $ on top of a GaP resonant nanoantennas with different radii (in red), and on top of the bare SiO$ _2 $ substrate (in blue). The dashed line marks the unstrained exciton energy. Inset: experimental PL enhancement factor, $ \braket{EF} $, extracted from the integrated PL intensity. (c) Time resolved PL traces from a WS$ _2 $ monolayer coupled to a GaP nanoantenna with r = 90 nm (in red) and on the glass substrate (in blue), revealing a two-fold reduction of the exciton decay dynamics ascribed to the Purcell effect. The data is fitted with a single exponential decay (solid lines).}
	\label{fig:Fig_rad}
\end{figure*}

\noindent
\textbf{Radiative rate enhancement in coupled WS$_2$ excitons}
We further investigate the PL properties of coupled WS$ _2 $ excitons by means of steady state and time resolved optical spectroscopy.
Figure \ref{fig:Fig_rad}a shows the PL map of the monolayer transferred on top of a resonant GaP nanoantennas array. The sample is excited with a 530 nm, 140 fs pulsed laser, with repetition rate of 80 MHz and average power of 14 nW. We scan the sample with piezoelectric stages, and record the PL intensity with an avalanche photodetector (see Methods). We observe more than one order of magnitude enhancement of PL emission when the monolayer is placed on top of the nanoantennas, owing to the interplay of enhanced light emission and absorption rates of the coupled nanophotonic system \cite{Sortino2019}. Figure \ref{fig:Fig_rad}b shows the spectra of WS$ _2 $ on SiO$ _2 $ substrate and on GaP nanoantennas with varying radius and fixed height of 100 nm. Here, the PL is sent to a monochromator and CCD camera, where a tenfold increase in the integrated PL intensity for coupled WS$ _2 $ is observed. A maximum of PL is found for the nanoantenna with radius of approximately 90 nm, as expected from the optimized spectral overlap between Mie modes and WS$ _2 $ excitons (Figure \ref{fig:figure2dantenna}a). To fully capture the effect  our hybrid nanophotonic platform we calculated the PL enhancement factor \cite{Koenderink2017}, $ \braket{EF} $, resulting in values exceeding 200 (see Inset Figure \ref{fig:Fig_rad}b and Supplementary Note III). Moreover, the PL peaks exhibit a redshift for coupled monolayers, consistent with the occurrence of tensile strain at the edges of the nanoantenna \cite{Sortino2020}. We found a maximum red shift of 21 meV, compared to the monolayers on flat substrate, corresponding to 0.4\% tensile strain \cite{Niehues2018a}.

We then studied the PL dynamics in coupled and uncoupled monolayer by collecting time resolved luminescence traces of with a streak camera setup (see Methods). As depicted in Figure \ref{fig:Fig_rad}c, we observe a two-fold reduction of the decay lifetime for monolayers coupled to the nanoantenna near field, showing lifetimes of $( 128 \pm 1 )$ ps, compared to $ (298 \pm 6) $ ps on bare substrate. 
Note, PL lifetimes in TMDCs are mainly limited by non-radiative processes, even at low power densities \cite{Wang2017}, hindering the extraction of an effective value of the spontaneous rate enhancement. In our experiments, we employed a pump fluence of 120 $ \mu $J/cm$ ^2 $ to obtain appreciable signal to noise ratio, too large to neglect the impact of non-radiative processes in time resolved experiments \cite{Sun2014}.
To elucidate the role of strain in the PL experiments, we prepared a control sample, where a WS$ _2 $ monolayer is transferred on top of SiO$ _2 $ nanopillars with the same geometry and dimensions of the GaP nanoantennas. The nanopillar provides a deformation centre, where strain is introduced in the monolayer, while lacking optical Mie resonances owing to its lower refractive index \cite{Sortino2021a}. The experimental analysis of the control sample is presented in Supplementary Note IV. In strained WS$ _2 $ we observe the presence of a strained exciton peak at the pillar sites, where a slight enhancement of the PL intensity could be detected. However, we observe no changes in the PL lifetime when compared to unstrained monolayers (Supplementary Figure 5b). This effect is consistent with previous reports where strain does not significantly impact the luminescence lifetime of WS$ _2  $ monolayers \cite{Niehues2018a}, we thus ascribe the reduction of PL lifetime in our hybrid 2Ds semiconductor-dielectric system to the Purcell effect. 

\begin{figure*}[t]
	\centering
	\includegraphics[width=1\linewidth]{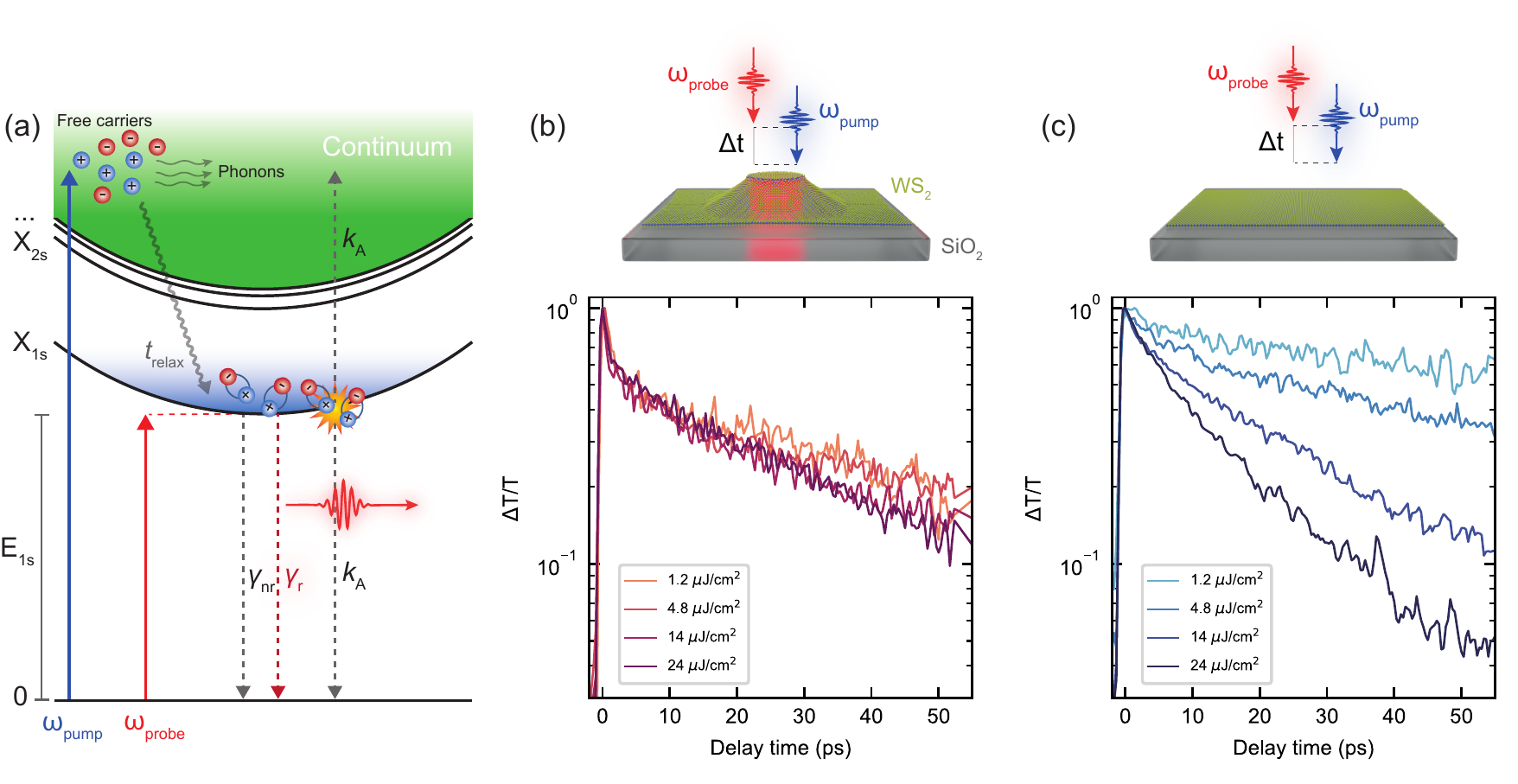}
	\caption{\textbf{Ultrafast dynamics and exciton-exciton annihilation processes in coupled and uncoupled WS$ _2 $ monolayers} (a) Schematics of the exciton formation and recombination dynamics under non-resonant excitation. A pulsed laser injects high energetic electrons and holes which thermalize by losing their kinetic energy via exchange to phonons and scattering, relaxing to the \textit{1s}-exciton state (X$ _{1s} $). The resulting exciton population recombines down to the ground state via radiative ($ \gamma_r $) and non-radiative ($\gamma_{nr}$) recombination processes, or undergo diffusion and scattering with other excitons resulting in exciton-exciton annihilation ($ k_A $). (b) Ultrafast differential transmittance ($ \Delta $T/T) of the WS$ _2 $ exciton when coupled to the resonant GaP nanoantenna (radius 90 nm, height 100 nm), acquired for different pump beam fluences. (c)  Fluence dependence of the differential transmittance of WS$ _2 $ exciton on SiO$ _2 $ substrate. The dynamics of uncoupled excitons exhibit the onset of EEA processes at fluences above  10 $ \mu $J/cm$ ^2 $, leading to a fast recombination of the exciton population compared to coupled excitons.}
	\label{fig:figureeea}
\end{figure*}

\noindent
\textbf{Ultrafast dynamics of coupled and uncoupled WS$_2$ excitons}
To obtain additional insight on the exciton recombination dynamics in hybrid 2D semiconductor-dielectric nanoantennas, we investigated the coupled nanophotonic systems by means of non-degenerate ultrafast pump-probe spectroscopy (see Methods and Supplementary Note V).
Figure \ref{fig:figureeea}a depicts a schematic of the excitons' dynamics in a monolayer TMDC under non-resonant optical excitation. The absorption of high energy photons leads to the formation of free carriers in the high-lying bands, which undergo relaxation and formation of the exciton species via emission of phonons on a sub-100 fs timescale \cite{Trovatello2020}. 
Due to the Wannier-Mott character, excitons diffuse in the crystal and the overall dynamics are described as the product of the spontaneous emission rate ($ \gamma_r $) and the non-radiative recombination rate ($\gamma_{nr}$), e.g. from phonon, defects or carrier scattering, and EEA.
The exciton population dynamics are described by the following equation:
\begin{equation}
	\frac{dN}{dt}=G + D\nabla^2N-(\gamma_r + \gamma_{nr}) N -k_A N^2
\end{equation}
where $ G $ is the generation rate of excitons, $ D $ is the diffusion coefficient and $k_A$ is the EEA coefficient. When the exciton population density ($N$) is small, the recombination dynamics are dominated by the sum of radiative and non-radiative processes. As $ N $ increases, the quadratic term of the EEA starts to dominate the dynamics, leading to a fast decay of the photogenerated excitons \cite{Sun2014}. Diffusion related effects are be expected to take place on a fast timescale, owing to exciton in-plane diffusion coefficients in the order of 200 nm$ ^2 $/ps \cite{Mouri2014}, and further increased by the effect of strain-induced exciton funnelling \cite{Rosati2021}.

We compare the dynamics for excitons coupled with a resonant nanoantenna to that of a monolayer on bare SiO$ _2 $ substrate. 
By resonantly probing the \textit{1s}-exciton state, we investigate the impact of EEA processes in the exciton recombination dynamics as a function of the pump fluence, directly proportional to the photoexcited exciton population. 
Figure \ref{fig:figureeea}b shows the transient absorption dynamics for a WS$ _2 $ monolayer deposited on a resonant GaP nanoantenna (radius 90 nm, height 100 nm), as a function of the pump beam fluence. 
We observe that the pump has negligible effects on the exciton lifetime, exhibiting minor changes for the range of fluences employed in our experiments, at the same time presenting a fast recombination process at zero-time delay, in the order of 1 ps, independent on the excitation fluence. Remarkably, we observe the absence of the onset of EEA processes, as expected for uncoupled excitons under fluences above 10 $ \mu $J/cm$ ^2 $ \cite{Sun2014}, further confirmed in WS$ _2 $ deposited on the other resonant nanoantennas (see Supplementary Note VI). Note, that while the strain values extracted from the exciton redshift changes between nanoantennas, in the range 0.15 - 0.45 \%, the dynamics of coupled excitons are not significantly affected. These dynamics found for coupled excitons are in striking contrast with those observed in uncoupled monolayers. As shown in Figure \ref{fig:figureeea}c, in the transient absorption signal of a monolayer WS$ _2 $ on SiO$ _2 $ substrate we observe clear changes in the exciton dynamics as a function of the pump fluence, in the form of the onset of a bimolecular recombination process, as expected from the role of EEA dominating the dynamics even under fluences as small as 10 $ \mu $J/cm$ ^2 $ \cite{Sun2014}.
As EEA processes can be neglected at the lowest fluence, the observed reduction of the lifetime in coupled excitons can be interpreted as the main effect of the Purcell effect, increasing the radiative recombination rate and reducing the overall exciton lifetime. In Supplementary Note VII, we compare the exciton dynamics at longer timescales, where the population of uncoupled excitons decays in $\approx$100 ps for excitation above 10 $ \mu $J/cm$ ^2 $.

We demonstrate the role of optical Mie resonances on WS$ _2 $ excitons' dynamics in Supplementary Note VIII, where we transferred WS$_2$ monolayers on top of non-resonant GaP nanoantennas. As the Mie resonances are spectrally decoupled to the WS$ _2 $ exciton (Supplementary Figure 10d), we observe negligible PL enhancement and the presence of an EEA onset in the ultrafast dynamics, as observed for monolayers on SiO$ _2 $ substrates. Moreover, as the off-resonant antenna provides a larger contact area between GaP and WS$_2$, compared to smaller resonant antennas, we conclude that the dielectric permittivity of the surrounding material does not impact the exciton dynamics within the investigated size ranges \cite{Steinhoff2021}.

 \begin{figure*}[t]
 	\centering
 	\includegraphics[width=1\linewidth]{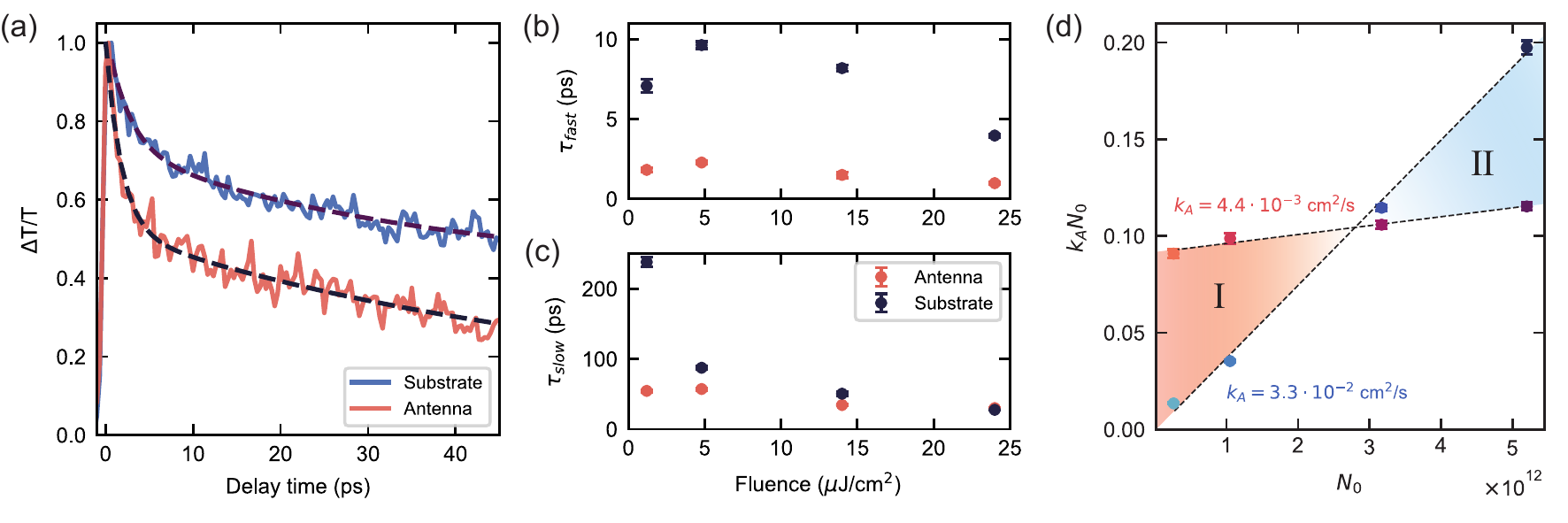}
 	\caption{\textbf{Suppression of exciton-exciton annihilation via radiative rate enhancement} (a) Differential transmittance traces for WS$ _2 $ on substrate (in blue) and on resonant GaP nanoantenna (in orange), excited with a pump fluence of 1.2 $ \mu $W/cm$ ^2 $. The data is fitted with a biexponential model (dashed lines). (b-c) Values of the fast, $ \tau_{fast} $, (b) and slow, $ \tau_{slow} $, (c) time component of the biexponential fit. The fast component is mostly unaffected by the pump fluence. We extract an enhancement of the absorption rate value of $\sigma = \tau_{fast}^{off}/\tau_{fast}^{on}= 3.9$ at low pump fluences. For the fast component, at the lowest fluence we can exclude the effect of EEA processes and extract an enhancement of the spontaneous emission rate, or Purcell factor, equal to $ F_P=\tau_{slow}^{off}/\tau_{slow}^{on}=4.5$. (d-e) Values of the EEA coefficient ($k_A$) extracted from the differential transmittance values for WS$ _2 $ coupled to a resonant nanoantenna (d) and on SiO$ _2 $ substrate (e), exhibiting a reduction of one order of magnitude for WS$ _2 $ coupled to resonant nanoantennas.}
 	\label{fig:Figure_Purcellultrafast}
 \end{figure*}

\noindent
\textbf{Radiative suppression of exciton-exciton annihilation via enhanced light-matter coupling}
The ultrafast optical response of 2D semiconductors is described by the interplay of different effects determining their dynamics \cite{DalConte2020}. For instance, band gap renormalization and changes in the binding energy, resulting in spectral shifts of the exciton resonance, or broadening of the exciton linewidth, via scattering and collisions, competing with Pauli blocking of photoexcited carriers. However, for excitation energies above the WS$ _2 $ bandgap, as in our experiments, the ultrafast response is dominated by Pauli blocking \cite{Trovatello2022a}. Furthermore, large exciton binding energies hinder auto-ionization of resonantly excited excitons in high-lying states, and we exclude thermal effects in our experiments following a linear dependence of the differential transmission signal modulation at zero time delay as a function of the pump power (see Supplementary Note IX). In our experiments, the dynamics are thus described within the picture of a thermalized population of the WS$ _2 $ \textit{1s}-exciton state.
We fit the ultrafast exciton dynamics with a biexponential model (Figure \ref{fig:Figure_Purcellultrafast}a) and extract the values of the fast and slow lifetime components, as shown in Figure \ref{fig:Figure_Purcellultrafast}b.
For all the coupled monolayers, we observe the presence of a fast sub-ps decay of the exciton population, independent on the pump fluence, and limited by the temporal resolution of our setup. As shown in Supplementary Figure 6, we compare the ultrafast dynamics of strained excitons to that of flat excitons on SiO$ _2 $ substrate, where no sub-ps dynamics are observed in strained WS$ _2 $. 
This fast component is related to the formation process of excitons under above bandgap excitation \cite{Cunningham2016,Ceballos2017a} and proportional to the exciton population generated in the semiconducting layer, where excitons rapidly reach the thermalized state \cite{Shah1996,Estrada-Real2022}. As the fast dynamic is observed only for coupled excitons (see Supplementary Note VI), we ascribe this effect to the enhanced absorption rate via near field coupling with the optical Mie resonances, leading to higher exciton densities. We extract the near field induced absorption enhancement as the ratio of the slow lifetime components on the antenna ($ \tau_{fast}^{on} $) and on the substrate ($ \tau_{fast}^{off} $). At fluences of 1.2 and 4.8 $ \mu $J/cm$ ^2 $, we extract a constant value of $\sigma = \tau_{fast}^{off}/\tau_{fast}^{on}= 3.9$, while under higher fluences above 10 $\mu $J/cm$ ^2 $, the ratio is reduced by the impact of EEA processes. As $\sigma$ is proportional to the near field intensity, $\sigma\propto(|E|/|E_0|)^{2}$ \cite{Koenderink2017}, the extracted value is consistent with the expected near field enhancement, as shown in Figure \ref{fig:figure2dantenna}d. 
At low fluences, the effect of EEA and non-radiative processes can be neglected, allowing the extraction of a lower bound value of the enhancement of the spontaneous emission rate for coupled excitons. By comparing the slow lifetime components ratio at low power, we extract a Purcell factor of $ F_P=\tau_{slow}^{off}/\tau_{slow}^{on}=4.5$.
Finally, we extract the EEA coefficient, $ k_A $, by fitting the data with a rate equation model (see Supplementary Note X for details on the model and fitting procedure) and plot the values as a function of the exciton density in Figure \ref{fig:Figure_Purcellultrafast}d.
From the slope of the linear fit, we extract a value of $ k_A = (4.4 \pm 0.6) \times 10^{-3}$ cm$^2$/s in the case of coupled WS$ _{2} $ monolayer. Instead, for uncoupled monolayer on a glass substrate we found $ k_A = (3.3 \pm 0.3)\times 10^{-2}$ cm$^2$/s, one order of magnitude higher than for the nanoantenna coupled WS$ _2 $ and consistent with previous reports \cite{Yuan2015b}. 

Comparing the number of scattering events  ($k_AN_0$) to the injected exciton density ($ N_0 $), we observed two regimes determined by the EEA processes, denoted as Region I and II in Figure \ref{fig:Figure_Purcellultrafast}d.
In Region I, at lower exciton densities, coupled excitons exhibit a higher number of scattering events compared to the uncoupled case. This is a direct consequence of the enhanced absorption rate, resulting in a larger $ N_0 $ at the antenna position, directly inducing the fast picosecond recombination observed in our experiments. 
However, above a certain density threshold, highlighted as Region II in Figure \ref{fig:Figure_Purcellultrafast}d, we observe that $k_AN_0$ values for uncoupled excitons overtake the value for coupled ones, which, on the contrary, exhibit only a slight increase. Here, the exciton recombination is dominated by the radiative decay rate, as more excitons are depleted faster via the enhanced radiative channel. The lower EEA coefficient in coupled excitons is a direct consequence of the quadratic dependence of $ k_A $ to the excitonic population shown in Equation 2. The probability for excitons to diffuse and participate in scattering processes is hindered by the fast radiative recombination, reducing the overall number of non-radiative EEA events, directly increasing the quantum efficiency and realizing higher exciton densities in coupled WS$ _{2} $ monolayers.

\noindent
\textbf{Discussion}

\noindent
In summary, we demonstrated the suppression of EEA processes in a hybrid 2D semiconductor-dielectric system by coupling excitons in monolayer WS$ _2 $ with optical Mie resonances of GaP dielectric nanoantennas. The system reaches an intermediate light-matter coupling regime with PL enhancement factors above 10$ ^2 $, as a result of enhanced absorption and spontaneous emission rates. From the ultrafast exciton dynamics, we show the resilience of nanoantenna coupled excitons to sustain higher pump fluences without the onset of EEA processes.
We extract one order of magnitude smaller EEA coefficients, together with absorption enhancement of 3.9 and Purcell factor of 4.5. Owing to the increased radiative recombination rate, the exciton population is depleted faster via photon emission, suppressing the onset of EEA observed in uncoupled TMDCs monolayers. 
Engineering the photonic environment represents a novel opportunity to further reduce EEA processes in 2D semiconductors, for instance via integration with van der Waals metasurfaces \cite{Kuhner2022b}. Moreover, rationally designed hybrid nanophotonic systems based on 2D materials offer a vast toolbox for shaping and controlling light field at the nanoscale, for realizing higher spontaneous emission rates.
Suppression of EEA via integration of 2D semiconductors with hybrid architectures can be extended to van der Waals heterostructures and Moiré systems, merging nanophotonic with many body physics and strongly correlated exciton phases \cite{Mak2022}, and to novel hybrid platforms based on 2D and bulk van der Waals materials, as building blocks of optically resonant nanostructures \cite{Lin2022,Weber2022}.

\bigskip

\noindent
\textbf{Methods}

\noindent
\textbf{Sample fabrication}
GaP nanoantennas are fabricated with a top-down process using a combination of electron beam lithography (EBL) and reactive ion etching (RIE).  
100 nm GaP and 80 nm SiO$ _2 $ are grown on a glass substrate by sputtering.
A double layer of Polymethylmethacrylate (PMMA) resist and an additional conducting layer, to avoid charging effects, are spin coated on the sample.
The resist is patterned with an EBL system, at 30 kV acceleration voltage, 20 $\mu$m aperture, dose of 330 $ \mu$C/cm$ ^2 $, and then developed in a 1:3 solution of Methylisobutylketone (MIBK) and isopropanol. 
3 nm Ti and 30 nm Au are evaporated with an electron beam evaporator, acting as etching mask. We first removed the SiO$ _2 $ layer via RIE and etch the gold with a standard gold etchant. The SiO$ _2 $ left on the sample is used as a mask for the GaP RIE step, and is finally removed with an additional RIE etching, resulting in the desired GaP nanostructures on fused silica substrate.
Monolayers of WS$ _2 $ are exfoliated from a commercial single crystal (HQ Graphene) and deposited on top of the nanoantennas with an all-dry transfer technique based on polydimethylsiloxane (PDMS) \cite{Castellanos-Gomez2014a} in a home build transfer setup.

\noindent
\textbf{Numerical simulations}
FDTD simulations were carried out with a commercial software (Ansys Lumerical). The refractive index of the amorphous GaP film are extracted from ellipsometry measurements \cite{Tilmann2020}.

\noindent
\textbf{Optical spectroscopy}
Dark field scattering experiments are performed in a commercial WiTec system with a broadband white light source. For PL and ultrafast spectroscopy experiments, we show a detailed schematics of the experimental setup in Supplementary Note V. 
The sample is mounted on a piezoelectric stage coupled to an inverted microscope system for mapping and fine tuning of its position. As excitation, we employ the frequency doubled output of a 180 femtoseconds pulsed tunable Ti:sapphire laser (Coherent Chameleon Ultra II) with repetition rate of 80 MHz. 
The PL signal is collected with a monochromator and CCD detector (Princeton Instruments) or an avalanche photodetector (MDP) for PL mapping.
The time resolved PL is acquired by directing the light to a monochromator and streak camera system (Hamamatsu). 
For pump-probe experiments, the frequency doubled Ti:sapphire laser is modulated at a frequency of 1990 Hz, with a mechanical chopper, and is used as the pump beam. The same laser drives a tunable optical parametric oscillator, which frequency doubled signal output is sent to an optical delay line and used as the probe.  We excite the 2D semiconductor with laser pulses at 435 nm (2.85 eV), while resonantly probing the \textit{1s}-exciton transition of WS$ _2 $ monolayers at approximately 620 nm (2 eV), tuned to the maximum modulation of the WS$ _2 $ response for each sample. The ultrafast dynamics are then recorded with a photodiode at the output of a grating monochromator, and with lock-in amplification. As the probe beam is resonantly exciting the exciton population, we carefully calibrated the impact of the probe energy by avoiding the presence of a fast decay peak in the response of the WS$ _2$. This is shown in Supplementary Note XI. We set the probe beam to a fluence of 1.5 $ \mu $J/cm$ ^2$ for all our experiments.

\bigskip
\noindent
\textbf{Data availability}

\noindent
The data that support the findings of this study are available from the corresponding author upon reasonable request.

\bigskip
\noindent
\textbf{Acknowledgements}

\noindent
S.A.M acknowledges the Lee Lucas chair in physics,  funding by the EPSRC (EP/WO1707511) and the Australian Research Council (Centre of Excellence in Future Low-Energy Electronics Technologies - CE 170100039). L.S. further acknowledges funding support through a Humboldt Research Fellowship from the Alexander von Humboldt Foundation. Our studies were  partially supported by the Center for NanoScience (CeNS) - Faculty of Physics, Ludwig-Maximilians University Munich. 

\noindent
\textbf{Conflict of interest}

\noindent
The authors declare no competing interests.

\clearpage

\appendix
\renewcommand{\figurename}{SUPPLEMENTARY FIGURE}
\setcounter{figure}{0}   
\linespread{1}

\section*{Supplementary information for: Radiative suppression of exciton-exciton annihilation in a two-dimensional semiconductor}

\newpage
\section*{Supplementary note I: Bright field and PL imaging of the hybrid 2D semiconductor-dielectric nanoantenna sample}

\begin{figure}[h]
	\centering
	\includegraphics[width=0.8\linewidth]{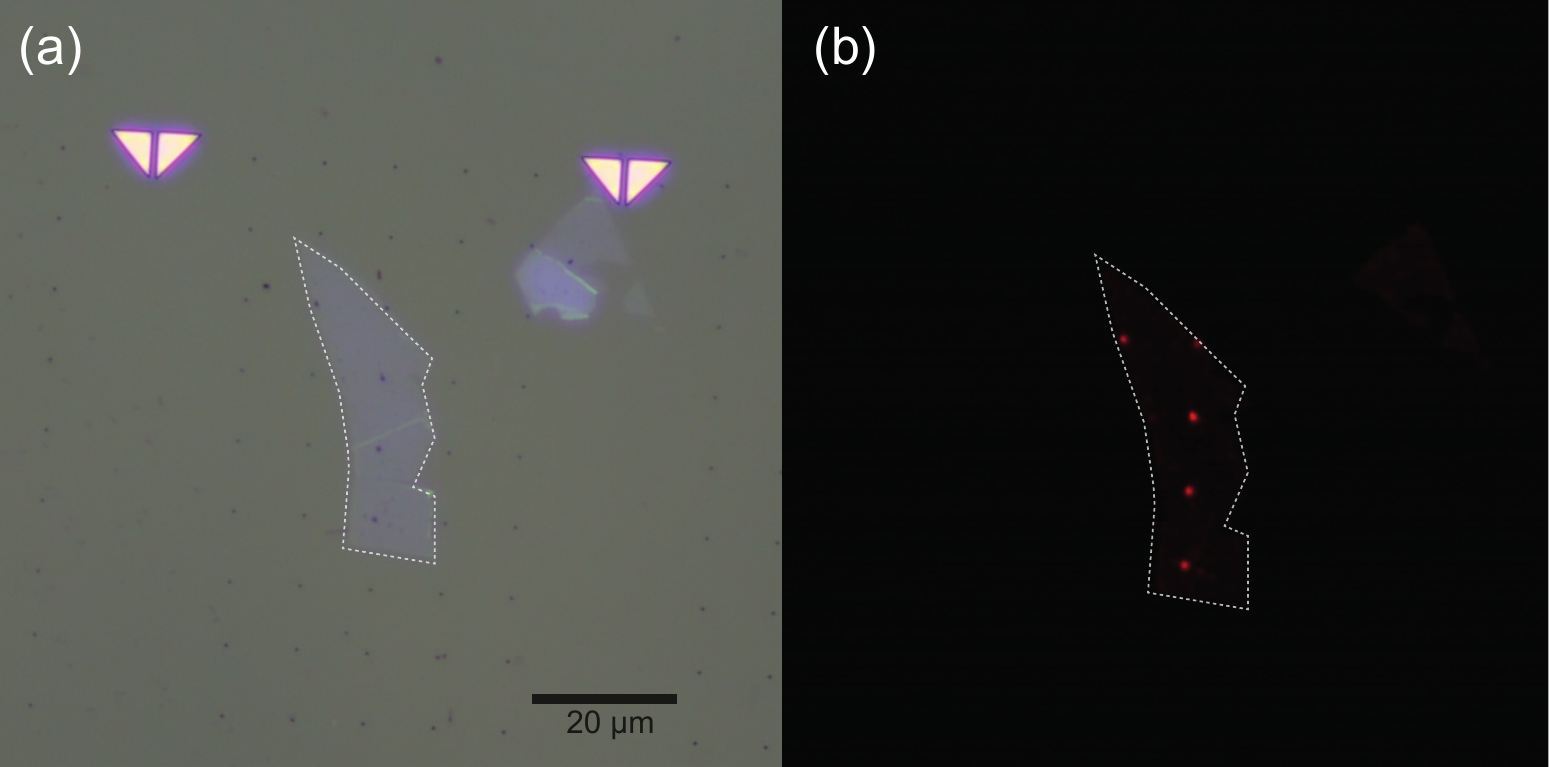}
	\caption{(a) Optical microscope bright field imaging of a WS$ _2 $ monolayer transferred on top of an array of GaP nanoantennas. The dashed white line outlines the transferred monolayer. (b) PL imaging of the same sample, acquired in the same microscope with the use of spectral filtering \cite{Alexeev2017}, showing the enhanced WS$_2$ PL intensity at the nanoantenna positions.}
	\label{fig:figs1}
\end{figure}

\section*{Supplementary note II: Dark field scattering of hybrid 2D semiconductor-dielectric nanoantennas}

\begin{figure}[h]
	\centering
	\includegraphics[width=0.55\linewidth]{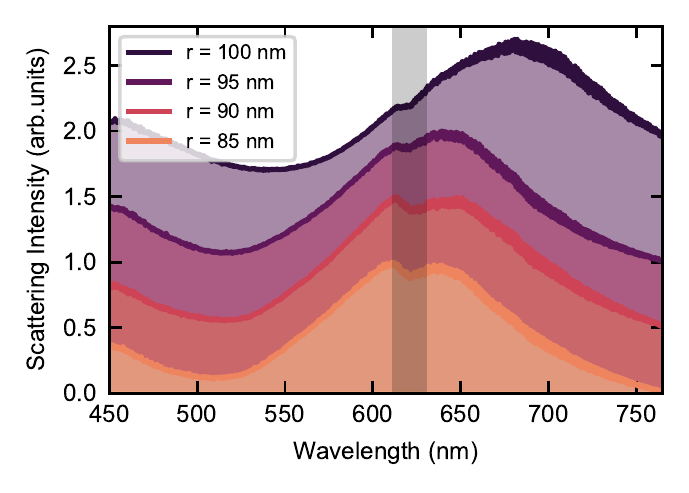}
	\caption{Dark field scattering spectra of GaP nanoantennas with WS$ _2 $ monolayer deposited on top. The spectra are shifted vertically for display purposes. The gray shaded area corresponds to the WS$ _2 $ exciton resonance.}
	\label{fig:figs2}
\end{figure}

\pagebreak
\section*{Supplementary note III: Photoluminescence enhancement factor}
\noindent
The PL enhancement factor, $\braket{EF}$, is extracted from experiments as \cite{Sortino2019}: 

\begin{equation}
	\braket{EF} = \dfrac{I_{on}}{A_{r}}\left(\dfrac{I_{off}}{A_{laser}}\right)^{-1}
	\label{Eq-experiment}
\end{equation}

\smallskip
\noindent
where, $I_{on}$ is the integrated PL intensity for WS$_2$ coupled to the nano-antennas, $I_{off}$ that of the uncoupled WS$_2$ on planar substrate. These are normalized on the PL excitation area, $A_{laser}$, defined by the excitation laser spot size, and the geometrical cross section of the cylindrical nanoantenna, $A_{r}=\pi r^2$.

\begin{figure}[b]
	\centering
	\includegraphics[width=1\linewidth]{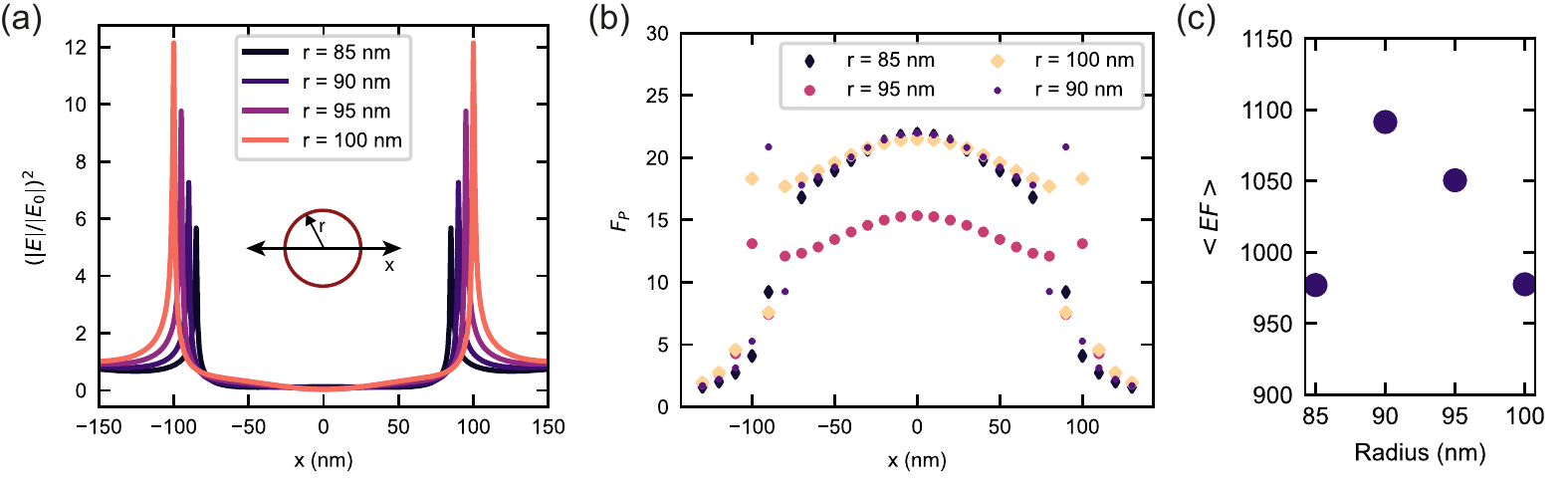}
	\caption{(a) Profile of the numerically FDTD simulated near field intensity, taken along the x-axis as shown in the inset. (b) Purcell factor for a dipole at different position on the x-axis, placed 0.5 nm above the nanoantenna surface. (c) Maximum values of the $\braket{EF}$ calculated from numerical simulations.}
	\label{fig:figs3}
\end{figure}

We calculated the upper bound of the $\braket{EF}$ from finite-difference time-domain (FDTD) numerical simulations, following:

\begin{equation}
	\langle EF\rangle \propto\dfrac{\sigma(\lambda_{exc})}{\sigma^{0}(\lambda_{exc})}\cdot \dfrac{F_P(\lambda_{em})}{F_P^{0}(\lambda_{em})}\cdot\dfrac{\eta(\lambda_{em})}{\eta^{0}(\lambda_{em})}
	\label{Eq-theory}
\end{equation}
\noindent
where $\sigma(\lambda_{exc})\propto(|E|/|E_0|)^{2}$ is the near field intensity, $ F_P(\lambda_{em}) $ the Purcell factor, and $ \eta(\lambda_{em}) $ the directivity enhancement of the nanoantenna coupled dipole. These values are normalized over the ones obtained for the uncoupled case of a dipole on silica substrate. We placed the dipole 0.5 nm above the edge of the nanoantenna, in order to maximize the values of field enhancement and $ F_P$ \cite{Sortino2019}.
Supplementary Figure \ref{fig:figs3}a shows the near field values at the surface of the nanoantennas, taken at the top surface along the x-axis, as shown in the figure inset. Supplementary Figure \ref{fig:figs3}b shows the values of the Purcell factor along the same x-axis, for an in-plane dipole emitting at 620 nm along the x-axis, placed 0.5 nm above the nanoantenna's surface. For directivity simulations, we calculated the power emitted by the dipole source in the far field, integrated over the solid angle defined by the objective numerical aperture. We found values of $ \eta/\eta_0 \approx 4 $ for nanoantennas of radius of 100 nm, increasing with the antenna radius up to a value of $ \eta/\eta_0 \approx 9 $ for nanoantennas with radius of 85 nm. The product of these three ratios gives the upper bound of the $\braket{EF}$, shown in Supplementary Figure \ref{fig:figs3}c.

\newpage
\pagebreak
\section*{Supplementary note IV: Strained WS$ _2 $ on S\lowercase{i}O$ _2 $ nanopillars}
\noindent
We fabricated silica nanopillars to eluidate the role of strain in the PL and dynamics in strained WS$ _2 $ monolayers.

\begin{figure}[h!]
	\centering
	\includegraphics[width=0.9\linewidth]{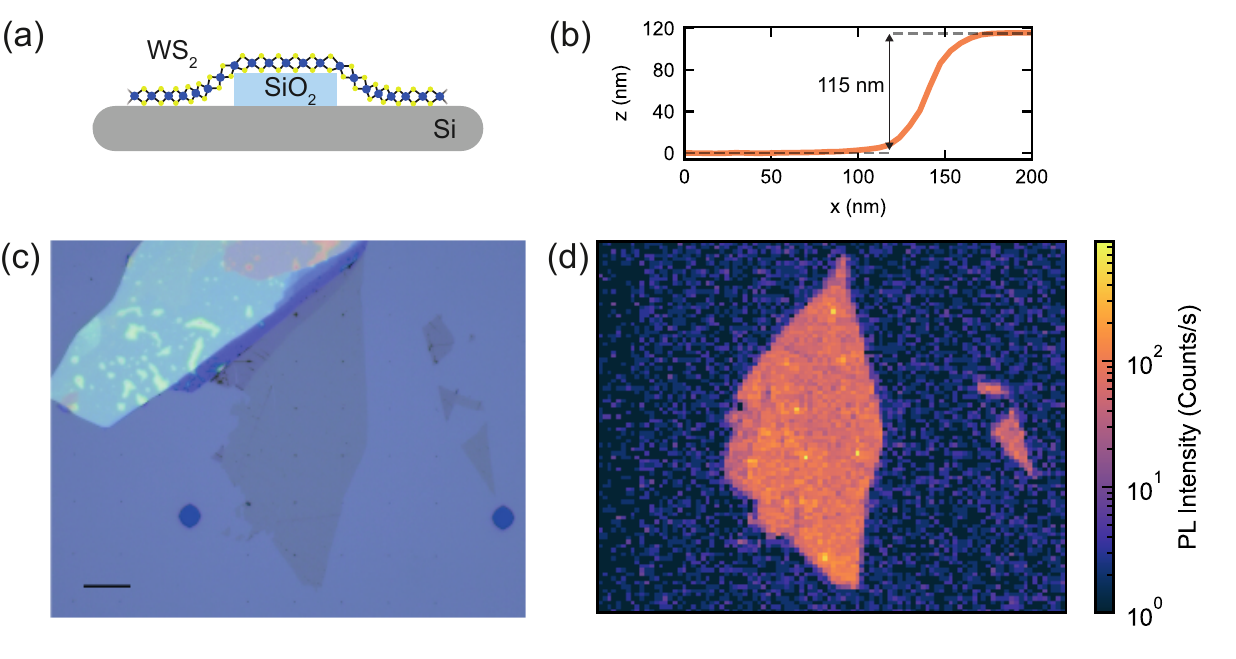}
	\caption{(a) Schematics of the silica nanopillar sample with monolayer WS$ _2 $. (b) AFM scan of a single nanopillar's egde. (c) Bright field image of the WS$ _2 $ on silica nanopillar sample. Scale bar 10 $ \mu $m. (d) Integrated PL intensity map of the silica nanopillar sample.}
	\label{fig:figs4}
\end{figure}

Supplementary Figure \ref{fig:figs4}a shows the schematic of the sample. The SiO$ _2 $ nanopillars are fabricated as follows. We employed a commercial Si wafer with a thermal SiO$ _2 $ layer of nominal thickness of 300 nm. We used a first etching step for thinning the oxide layer down to 100 nm, the target height of the nanopillars. We then use a combination of electron beam lithography and reactive ion etching (RIE) to pattern the silica nanopillars. 
Supplementary Figure \ref{fig:figs4}b shows an atomic force microscopy (AFM) profile of a fabricated nanopillar, revealing a height of 115 nm.
The WS$ _2 $ monolayer is exfoliated and transferred on the nanopillars with an all-dry PDMS based transfer technique. 
Supplementary Figure \ref{fig:figs4}c shows the bright field image of the final sample. In Supplementary Figure \ref{fig:figs4}d is shown the PL map of WS$ _2 $ excited with a 180 femtosecond laser at 532 nm and average power of 6.5 nW. We observe a small increase of the integrated PL intensity when the monolayer is on top of the nanopillars.

\begin{figure}[b]
	\centering
	\includegraphics[width=1\linewidth]{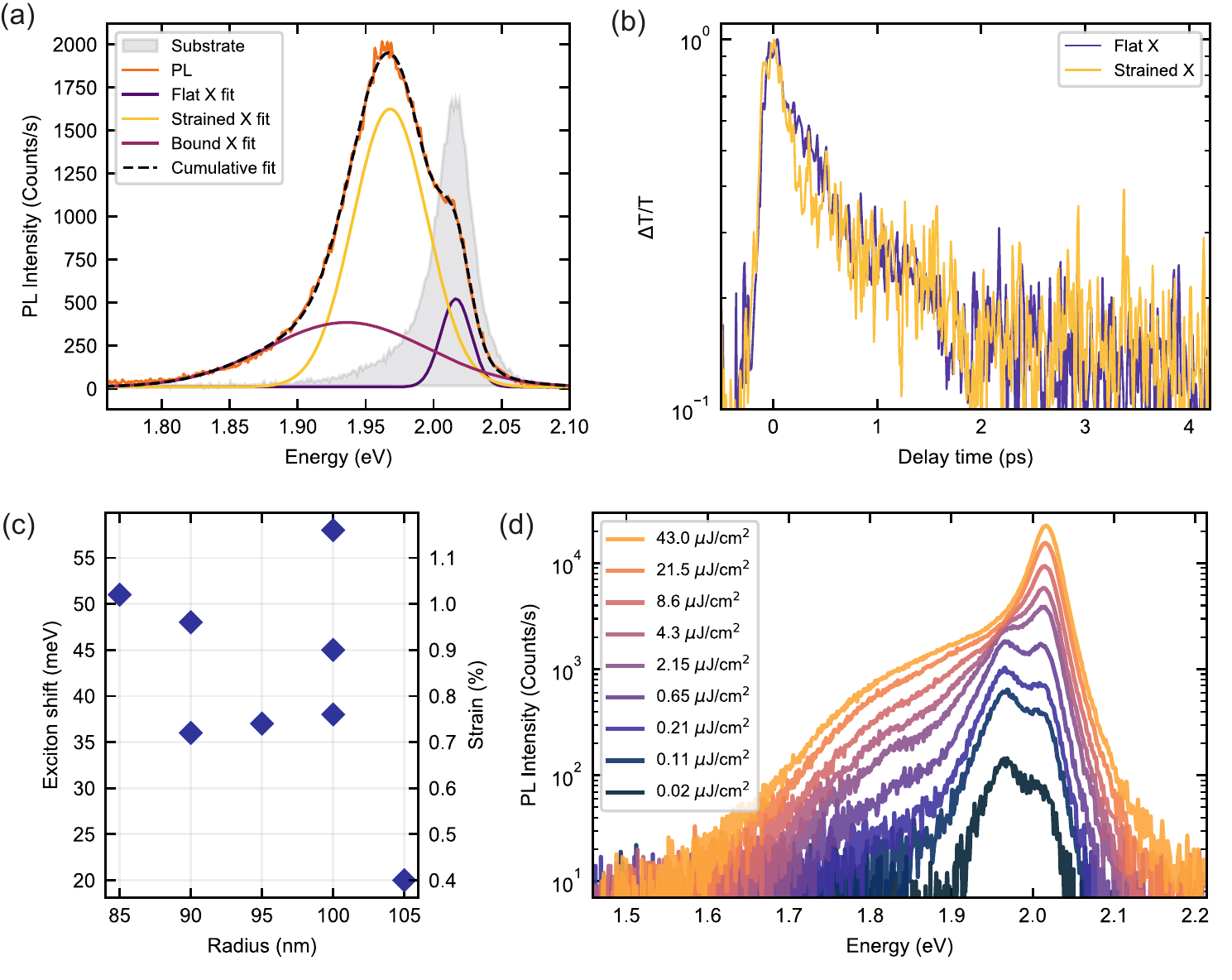}
	\caption{(a) Representative PL spectrum of WS$ _2 $ monolayer transferred on top of a silica nanopillar. The spectrum is fitted with Gaussian peaks, exhibiting an unstrained flat exciton (X) peak at high energy, a broad red shifted peak ascribed to local strain, and a defect bound exciton peak. In gray, the PL emission of WS$ _2 $ on bare silicon substrate. (b) PL dynamics traces for the flat exciton and strained exciton shown in Supplementary Figure \ref{fig:figs5}a. The traces are obtained with femtosecond pulsed excitation at 532 nm and 5 uW average power. (c) Strained exciton red shift, and corresponding tensile strain value, extracted from the fit of the PL spectra of WS$ _2 $ on silica nanopillars with different radii. (d) Power dependence of the WS$ _2 $ PL emission shown in Supplementary Figure \ref{fig:figs5}a.}
	\label{fig:figs5}
\end{figure}

\noindent
Supplementary Figure \ref{fig:figs5}a shows a representative PL spectrum collected on top of a single nanopillar, excited with a 532 nm laser pulse with average power of 1.4 $ \mu$W. At low power densities, we observe the presence of a broad excitonic peak red shifted from the exciton energy of the unstrained WS$ _2 $ on the substrate (in gray), related to the effect of tensile strain on the exciton energy \cite{Sortino2020}. We probed the PL dynamics and observed negligible differences in the PL lifetimes (see Supplementary Figure \ref{fig:figs5}b), as previously observed in strained WS$_2 $ \cite{Niehues2018a}. From the fit of the PL spectra of different nanopillars, we extract the energy shift of the strained exciton resonance and calculate the relative tensile strain value. This is shown in Supplementary Figure \ref{fig:figs5}c, where we extract values between 0.4\% and 1.1\%. Note, these are larger values than the ones observed in our hybrid 2D semiconductor-GaP nanoantenna sample.

Supplementary Figure \ref{fig:figs5}d shows the PL emission as a function of the excitation power for the same nanopillar shown in Supplementary Figure \ref{fig:figs5}a. At low power densities, the strained exciton peak intensity is comparable or higher than the signal from the surrounding unstrained excitons. We ascribe this effect to funnelling, which pushes bright excitons towards the strained area \cite{Rosati2021}. Moreover, at higher power densities the unstrained exciton peak dominates the PL emission, together with the appearance of a broad peak at energies below 1.9 eV.

\begin{figure}[h]
	\centering
	\includegraphics[width=0.7\linewidth]{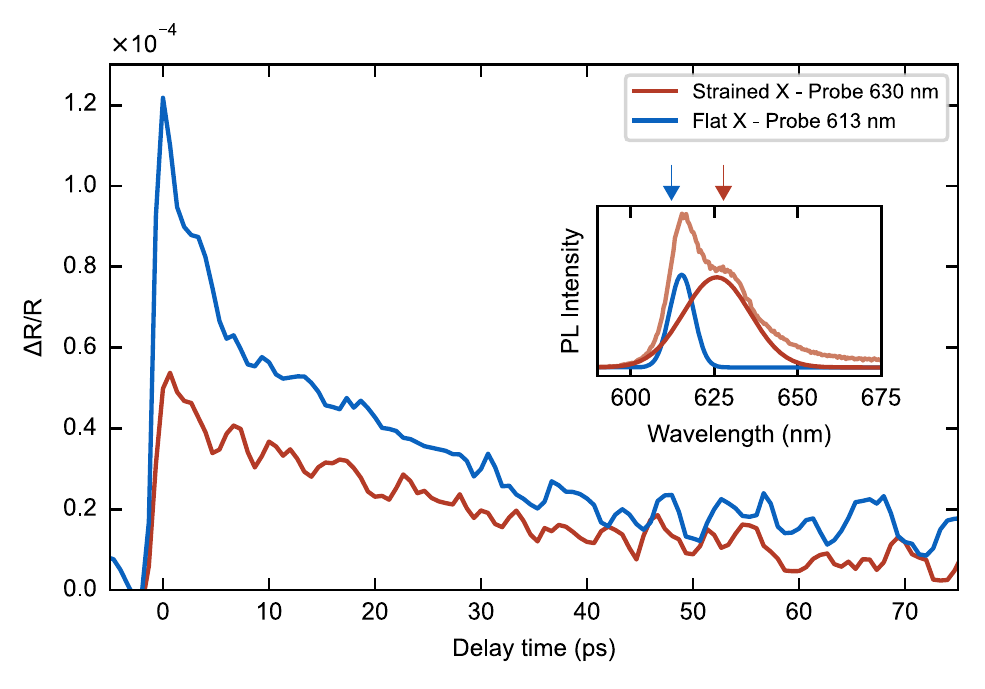}
	\caption{Ultrafast transient reflection ($ \Delta R/R $) for WS$ _2 $ excitons on silicon substrate (blue) and for strained excitons on silica nanopillar (red). Inset: PL spectrum of WS$ _2 $ on top of the silica nanopillar, the arrows indicate the probe wavelengths used in our experiments.}
	\label{fig:figs6}
\end{figure}

We further probed the ultrafast dynamics of flat and strained excitons. Supplementary Figure \ref{fig:figs6} shows the differential reflection ($ \Delta R/R $) for the WS$ _2 $ monolayer on top of a silica pillar with radius of 90 nm. We probed the exciton dynamics at two wavelengths, at 613 nm for the flat exciton and at 630 nm for the strained exciton (see PL in inset). The probe fluence was set at 1.2 $ \mu $W/cm$ ^2 $ and pump at 10.1 $ \mu $W/cm$ ^2 $. The high fluence conditions used to obtain a detectable signal in reflection measurements, are above the EEA onset. Indeed, for the flat exciton (blue curve) we observe a fast recombination dynamics related to EEA, while the strained exciton (red curve) we exhibit a slower recombination dynamics, with no fast component related to EEA. This exclude strain as the cause of the fast sub-ps dynamics observed when the WS$ _2 $ monolayers are coupled to GaP nanoantennas, as discussed in the main text.

\newpage
\pagebreak
\section*{Supplementary note V: Experimental setup schematics}

\begin{figure}[h]
	\centering
	\includegraphics[width=0.7\linewidth]{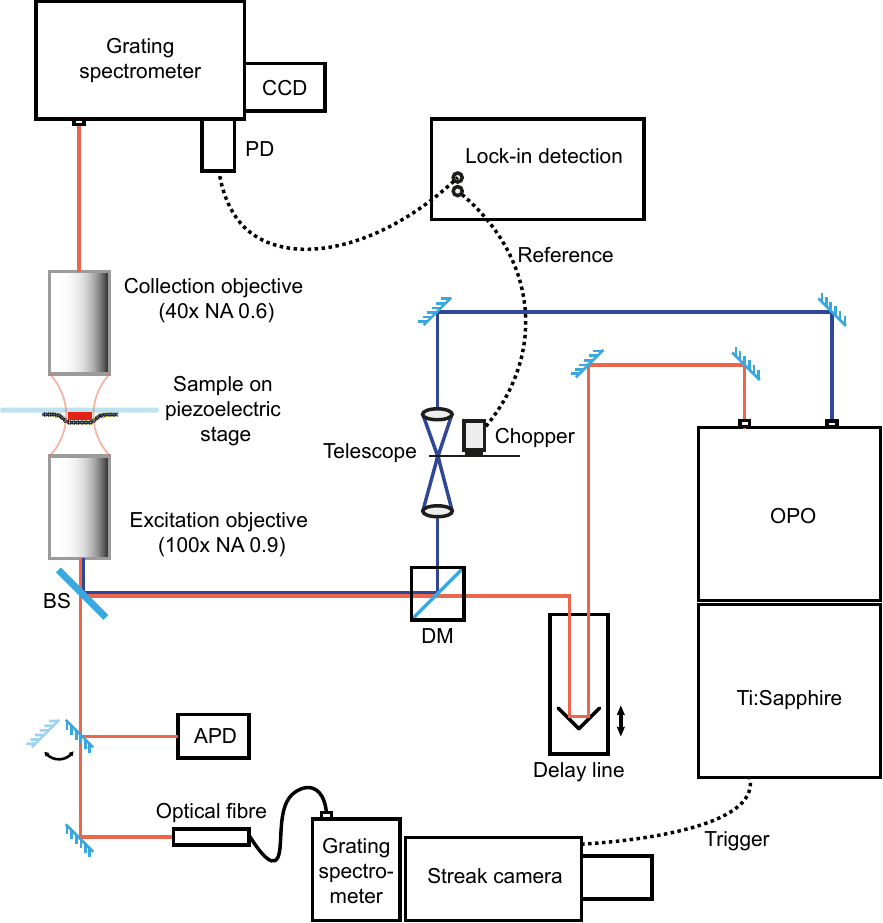}
	\caption{Schematics of the experimental setup. The second harmonic signal (SH) generated from the fundamental wavelength of a Ti:Sapphire laser is used as the pump pulse. The same laser pumps an optical parametric oscillator (OPO), the tunable SH of the OPO is used as probe beam. A telescope system is used to expand the pump beam for maximizing the spatial overlap with the focused probe beam on the sample. Before reaching the sample, the pump is modulated with a mechanical chopper, placed at the focal point of the telescope system. The probe beam is directed to an optical delay line. The two beams are recombined with a dichroic mirror (DM) and sent to an optical inverted microscope, after a beam splitter (BS), where the sample is sitting on piezoelectric stages, and excited with high numerical aperture (NA) objective and low NA collection objective. Transmitted light from the sample is spectrally filtered and sent to a grating spectrometer and CCD camera for PL detection, or to a photodiode (PD) for pump-probe experiments with lock-in detection. The reflected light from the sample is sent to either an avalanche photodiode (APD) for PL mapping, or to an optical fibre coupled with a grating spectrometer and streak camera for time resolved PL measurements.}
	\label{fig:figs7}
\end{figure}

\newpage
\pagebreak
\section*{Supplementary note VI: Ultrafast dynamics of WS$ _2 $ excitons coupled to additional resonant G\lowercase{a}P nanoantennas}

\begin{figure}[h]
	\centering
	\includegraphics[width=1\linewidth]{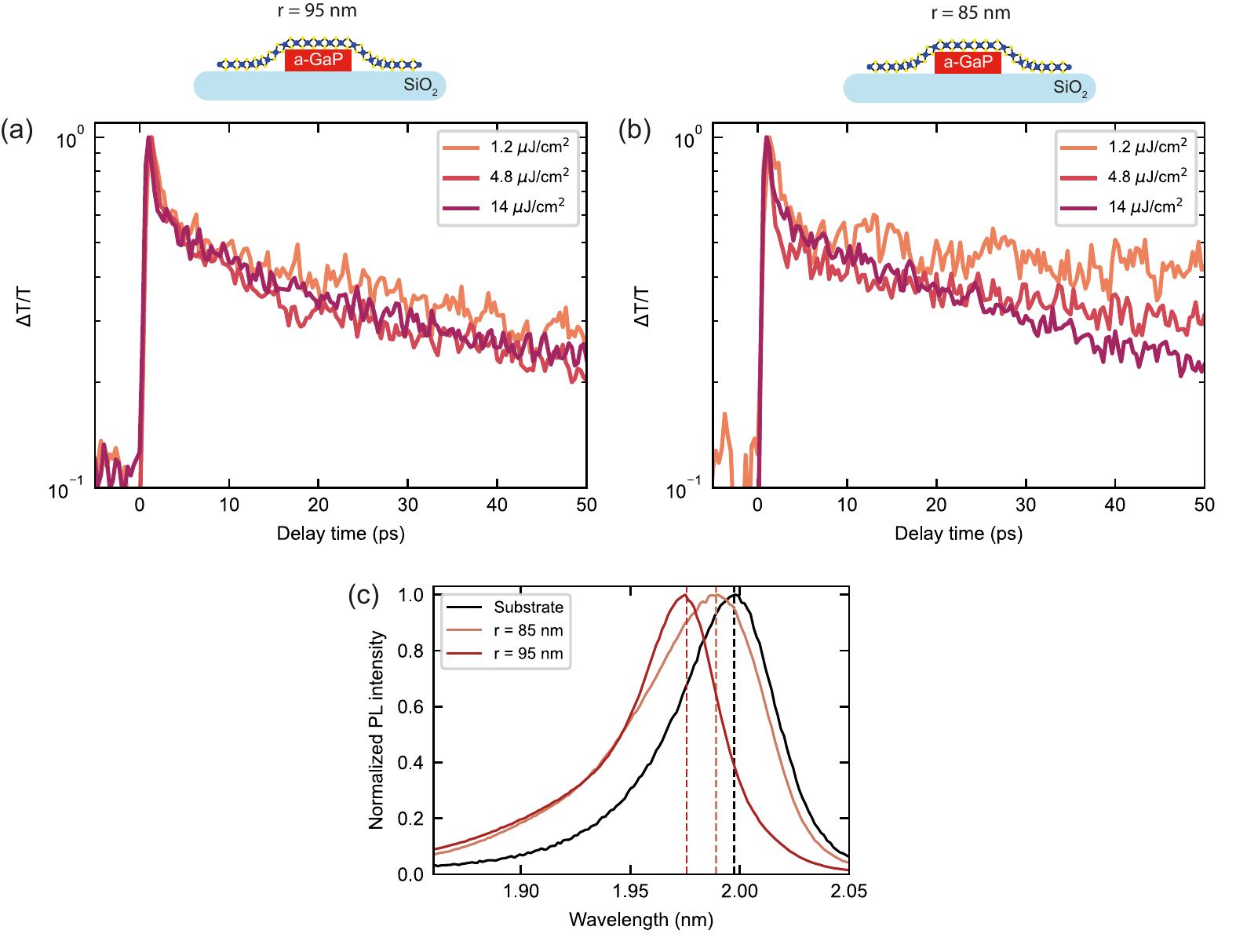}
	\caption{(a) Power dependence of the differential transmittance ($ \Delta $T/T) of the WS$ _2 $ excitonic state when coupled to a resonant antenna with radius of 95 nm. The extracted $ k_A = (6.0 \pm 1.7) \times 10^{-3}$ cm$^2$/s. (b) Power dependence of the differential transmittance ($ \Delta $T/T) of the WS$ _2 $ excitonic state when coupled to a resonant antenna with radius of 85 nm. The extracted $ k_A = (1.7 \pm 0.5) \times 10^{-2}$ cm$^2$/s. (c) Normalized WS$ _2 $ PL spectra on flat substrate (black) and on antennas with radius of 95 and 85 nm. We estimate the strain introduced in the monolayer by fitting the PL data and comparing the exciton spectral shift to the value of unstrained monolayer on bare substrate. We extract a value of tensile strain of 0.47\% for the WS$ _2 $ on the nanoantenna with radius of 95 nm, and of 0.16\% when deposited on the antenna with radius of 85 nm.}
	\label{fig:figs8}
\end{figure}

\newpage
\pagebreak
\section*{Supplementary note VII: Coupled and uncoupled exciton dynamics at long timescales}

\begin{figure}[h]
	\centering
	\includegraphics[width=1\linewidth]{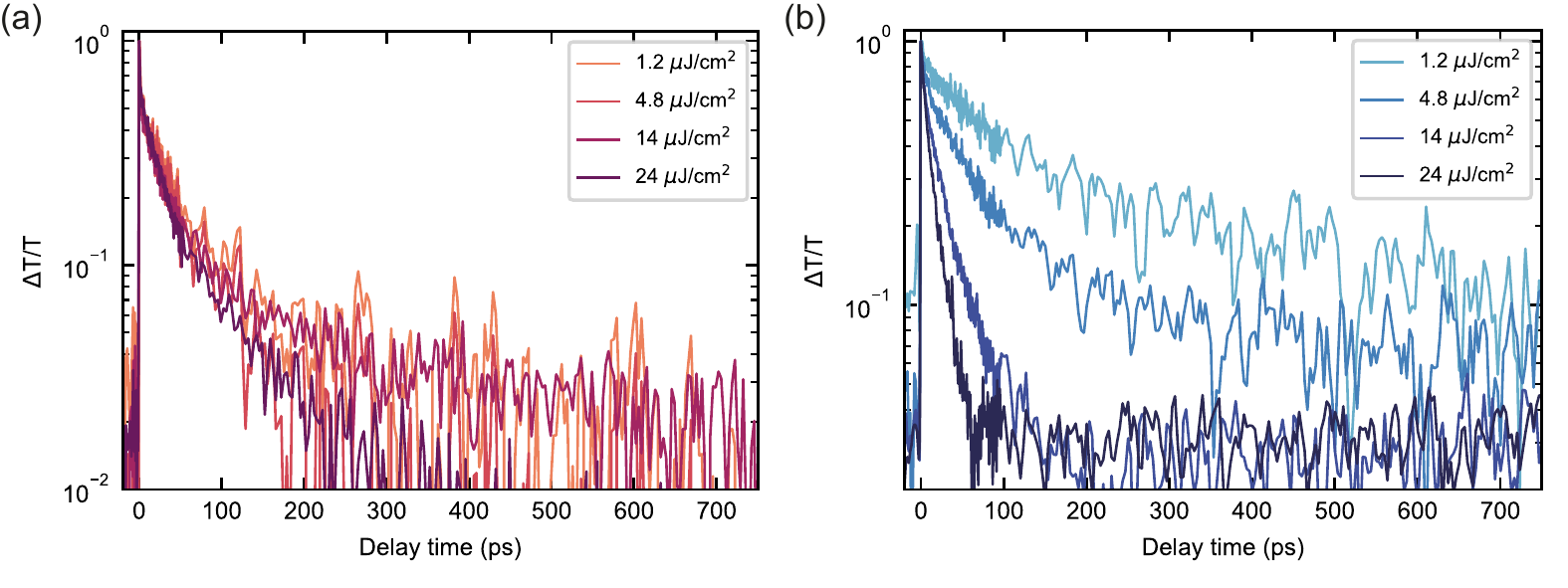}
	\caption{Dynamics of monolayer WS$_2$ excitons when coupled to a GaP nanoantenna (a) and on bare silica substrate (b), showing timescale dynamics longer than those presented in Figure 3 in the main text.}
	\label{fig:figs9}
\end{figure}

\newpage
\pagebreak

\section*{Supplementary note VIII: WS$_2$ monolayer coupled to non-resonant G\lowercase{a}P nanoantennas}

\noindent
We studied the effect of non-resonant GaP nanoantennas, with geometry of 50 nm height and radius of 320 nm, on the WS$ _2 $ PL  and dynamics.

\begin{figure}[b]
	\centering
	\includegraphics[width=0.9\linewidth]{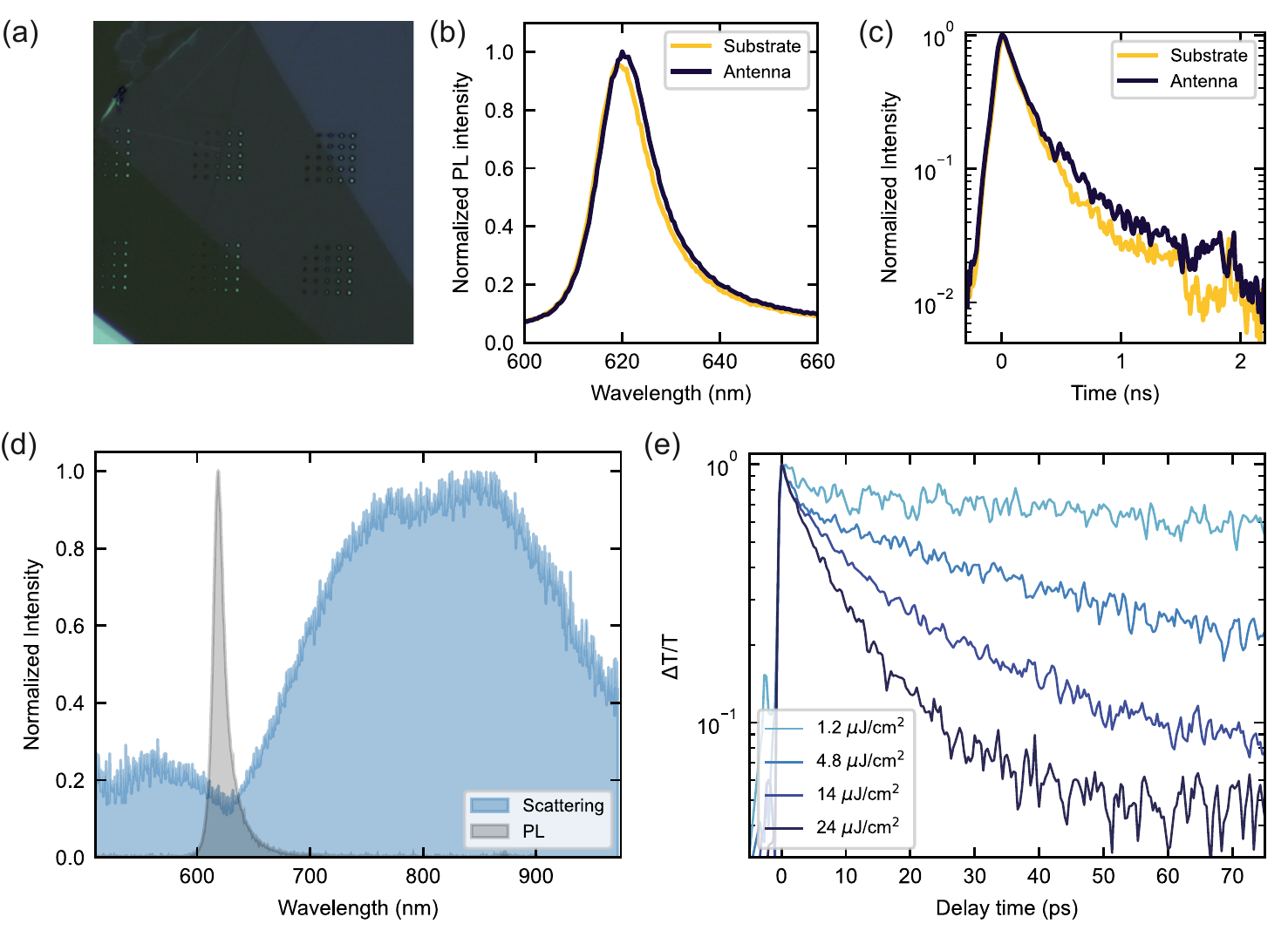}
	\caption{(a) Bright field image of WS$ _2 $ transferred on top of GaP nanoantennas. (b-c) WS$ _2 $ PL spectra (b) and PL dynamics (c), on silica substrate (black) and on non-resonant GaP nanoantenna (h= 50 nm, r= 320 nm). (d) Dark field scattering of a GaP nanoantenna (h= 50 nm, r= 320 nm) and PL of the transferred monolayer WS$ _2$ on top. (e) Ultrafast pump-probe exciton dynamics of WS$ _2$ on non-resonant GaP nanoantennas.}
	\label{fig:figs10}
\end{figure}

Supplementary Figure \ref{fig:figs10}a shows the bright field image of a transferred WS$ _2 $ monolayer on arrays of non-resonant GaP nanoantennas.  
As shown in Supplementary Figure \ref{fig:figs10}b, we observe negligible PL enhancement when the monolayer is transferred on top of such nanoantenna, and a small redshift in the exciton peak corresponding to 0.05\% strain. Supplementary Figure \ref{fig:figs10}c shows the PL dynamics, where a small increase of lifetimes at longer timescales is observed for monolayers on GaP. Supplementary Figure \ref{fig:figs10}d shows the dark field scattering of the studied sample, confirming the lack of spectral coupling between excitons and the optical resonances of the GaP nanoantenna, located at longer wavelengths above 700 nm.
In Supplementary Figure \ref{fig:figs10}e we report the exciton dynamics probed with the same experimental conditions as described in the main text. The effect of EEA is clearly observed in the dependence with the pump beam fluence. From the ultrafast dynamics, we extract a rate of $ k_A = (4.7 \pm 0.3)\times 10^{-2}$ cm$^2$/s, consistent with the values observed for unstrained WS$ _2 $ on silica substrates \cite{Yuan2015b}.

\section*{Supplementary note IX: Linear dependence of $ \Delta T/T $ at zero time delay}

\begin{figure}[h]
	\centering
	\includegraphics[width=1\linewidth]{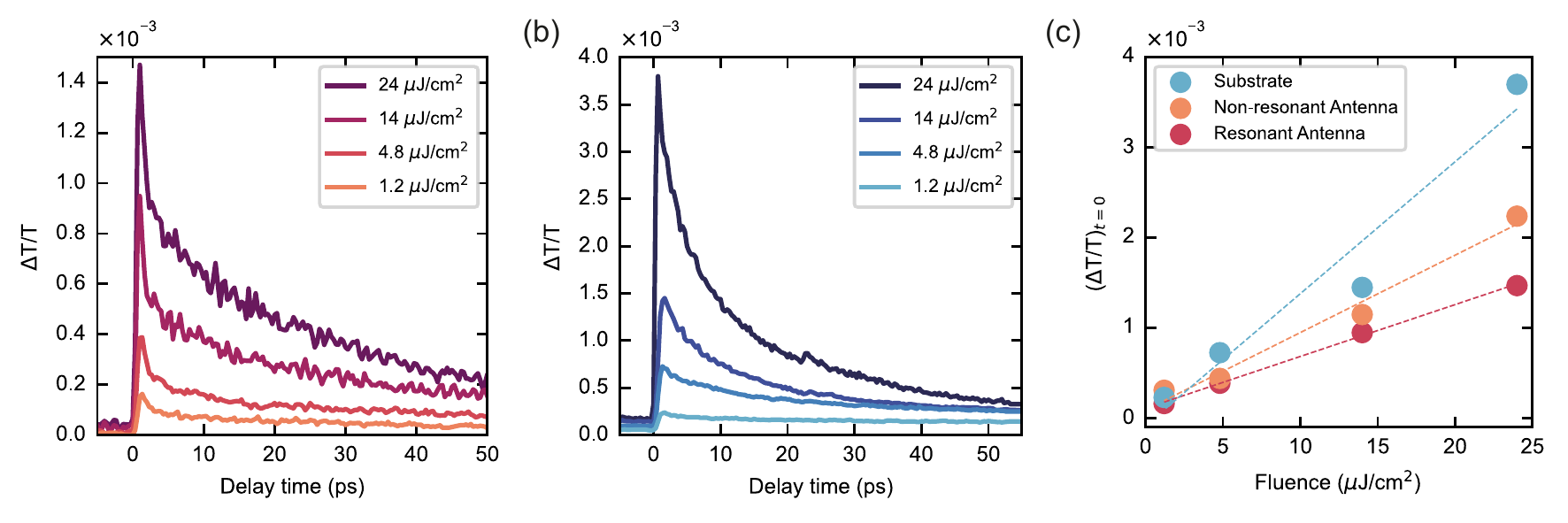}
	\caption{(a) Transient absorption, $ \Delta T/T $, of a WS$ _2 $ monolayer on resonant nanoantenna. (b) $ \Delta T/T $ of a WS$ _2 $ monolayer on silica substrate. (c) Dependence of $ \Delta T/T $ at zero time delay for WS$ _2 $ on resonant and on nonresonant nanoantennas, and on silica substrate. The lower modulation signal in the presence of the nanoantenna is related to the absorption and scattering of laser light by the nanostructure.}
	\label{fig:figs11}
\end{figure}

\newpage
\pagebreak
\section*{Supplementary note X: Fitting procedure to extract EEA rates}
\noindent
To extract the EEA coefficient ($ k_A $) from the experimental data, we implemented two procedures \cite{Sun2014,Kumar2014a}. 
The first procedure is done by fitting the data with the following equation:
\begin{equation}
	\frac{dN}{dt}=-k_A N^2
\end{equation}
This is a simplified case for Equation 2 in the main text. Here, we neglect the diffusion term and, as EEA processes are significant only at the early times, long recombination dynamics do not impact the extracted values of $k_A$ \cite{Kumar2014a}. 

The second procedure consists in normalizing the zero-time delay modulation of the transient absorption, $ (\Delta T/T)_{t=0} $, by the value of $ N_0$, calculated for an absorption of 0.1\% by WS$ _2 $ monolayer at the pump wavelength \cite{Li2014}.
As shown in Supplementary Figure \ref{fig:figs12}a, the data are then fitted with the following equation:
\begin{equation}
	\frac{N_0}{N(t)}-1=k_A N_0 t
\end{equation}
\noindent
From the slope of the linear fit, we extract the values of $ k_A N_0$, at different excitation powers. As it is proportional to $ N_0 $, we plot the obtained values as shown in Supplementary Figure \ref{fig:figs12}b. From the slope of the linear fit we then extract the value of $ k_A $, which we found is not significantly affected by the chosen procedure.

\begin{figure}[h]
	\centering
	\includegraphics[width=1\linewidth]{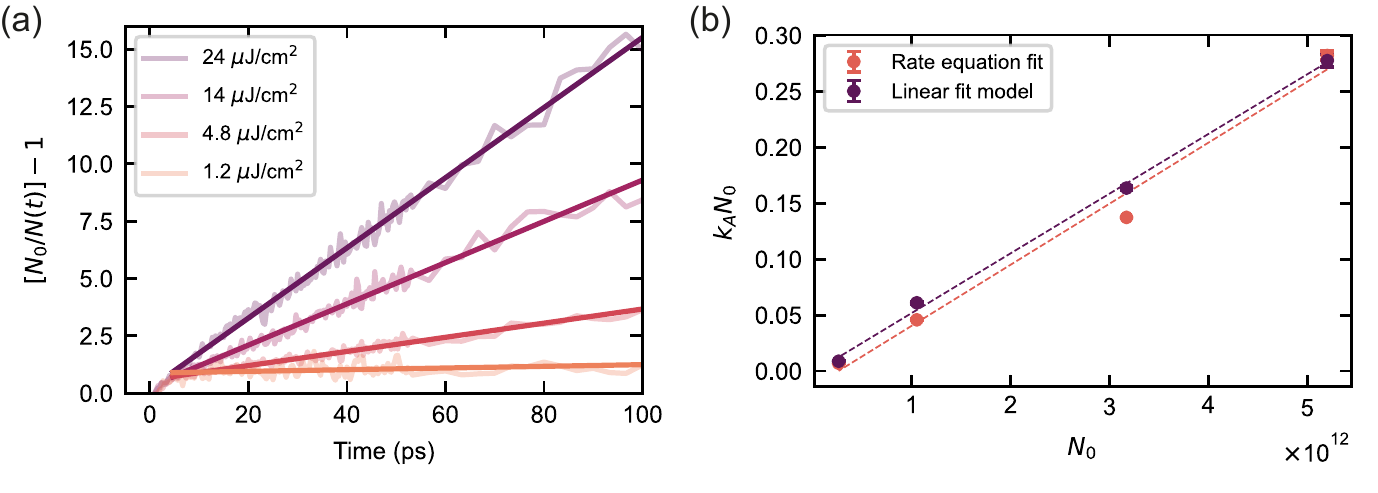}
	\caption{(a) Values of $[N_0/N(t)] - 1$ as a function of the time delay, and for different fluences, for a monolayer on silica substrate. From the slope of the linear fit we extract the $ k_A N_0 $ value. (b) Linear dependence of the $ k_A N_0 $ at different fluences, as a function of $ N_0 $. The slope of the linear fit gives the $ k_A $ value.}
	\label{fig:figs12}
\end{figure}

\newpage
\pagebreak
\section*{Supplementary note XI: Effect of resonant probe pulse fluence on exciton dynamics}
\begin{figure}[h]
	\centering
	\includegraphics[width=0.5\linewidth]{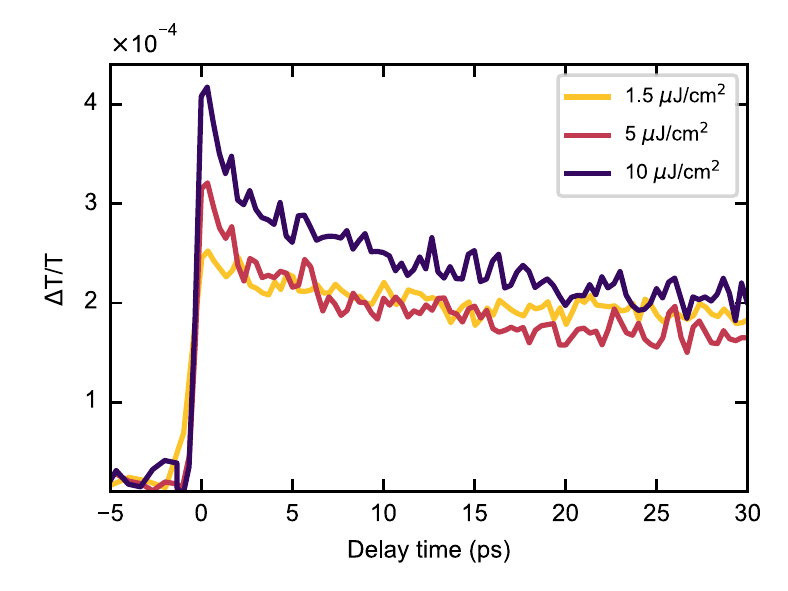}
	\caption{Effect of the resonant 618 nm probe beam fluence on the exciton dynamics of WS$ _2 $ on bare substrate. Here, the fluence of the pump beam at 435 nm is kept constant at a value of 2.4 $\mu$J/cm$^2$. As the resonant beam promotes excitons in the excited state, we calibrated the probe beam power until the disappearance of the fast decay dynamics at zero-time delay, in order to exclude the probe beam as a source of EEA in our experiments.}
	\label{fig:figs13}
\end{figure}

\end{document}